\documentclass{elsart}

\usepackage{natbib}

\usepackage{epsfig}

\usepackage{amssymb}

\begin{document}

\def \farcs{\hbox{$.\!\!^{\prime\prime}$}}
\def \lumstar{{${\rm L}^*_{\rm B}(z=0)=5.6\times10^9~h^{-2} {\rm L}_{\rm {B}\odot}~$}}

\begin{frontmatter}



\title{Current status of weak gravitational lensing}


\author[to1,to2]{Henk Hoekstra},
\author[to2]{H.K.C. Yee}, \and
\author[to2,carn]{Michael D. Gladders}

\address[to1]{CITA, University of Toronto, 60 St. George Street,\\
	M5S 3H8, Toronto Canada} 
\address[to2]{Dept. of Astronomy and Astrophysics, University of
	Toronto,\\ 60 St. George Street, M5S 3H8, Toronto Canada}
\address[carn]{Observatories of the Carnegie Institution of 
Washington,\\ 813 Santa Barbara Street, Pasadena, California 91101}

\begin{abstract}

Weak gravitational lensing of distant galaxies by foreground
structures has proven to be a powerful tool to study the mass
distribution in the universe. Nowadays, attention has shifted from
clusters of galaxies to the statistical properties of the large scale
structures and the halos of (field) galaxies. These applications have
become feasible with the advent of panoramic cameras on 4m class
telescopes.

In this review we will give an overview of recent advances in this
fast evolving field of astronomy. We start with a discussion of the
recent measurements of weak lensing by large scale structure (``cosmic
shear''), which can be used to constrain cosmological parameters.  We
also show how weak lensing can be used to measure the relation between
galaxies and dark matter (galaxy biasing) directly.

Another area that benefitted greatly from the current data is the weak
lensing by galaxies (galaxy-galaxy lensing). Weak lensing provides a
unique probe of the gravitational potential on large scales. Hence, in
the context of dark matter, it can provide constrains on the extent
and shapes of dark matter halos. Furthermore, it can test alternative
theories of gravity (without dark matter).  The best constraint comes
from the first detection of the anisotropy of the lensing signal
around lens galaxies, which suggest that the dark matter halos are
flattened. An isotropic signal, as predicted by MOND, is excluded
with 99\% confidence.

\end{abstract}

\begin{keyword}
dark matter \sep gravitational lensing


\end{keyword}

\end{frontmatter}

\section{Introduction}

Weak gravitational lensing is a valuable tool to study the (dark)
matter distribution of a range of objects in the universe. The
differential deflection of light rays by intervening structures allows
us to study the projected mass distribution of the deflectors, without
having to rely on assumptions about the state or nature of the
deflecting matter.

In 1984 Tyson et al. tried to detect the weak lensing signal induced
by an ensemble of galaxies using photographic plates. However, the
first succesful detection was made using CCD images of the massive
cluster of galaxies A1689 \citep{Tyson90}.  Although much theoretical
work concentrated on weak lensing by large scale structure
\citep[e.g.,][]{Miralda91,Blandford91,Kaiser92}, this area of astronomy
blossomed with study of the mass distribution in clusters of galaxies
\citep[e.g.,][for an extensive review
see Mellier 1999]{Bonnet94,Fahlman94,LK97,Hoekstra98,Hoekstra00} and galaxy
groups \citep{Hoekstra01a}.  These results were an important first
step in demonstrating the feasibility of the technique, but nowadays
more and more studies concentrate on the ``field''.

Adequate schemes to correct for a variety of observational distortions
(such as PSF anisotropy, seeing, camera distortion) were developed
during the 1990s \citep[e.g.,][]{KSB95,LK97,Hoekstra98,Kuijken99,
Kaiser00a,Bernstein02}. Together with the development of wide field
CCD cameras, these advances allowed for the first detections of weak
lensing by large scale structure in the spring of 2000
\citep{Bacon00,Kaiser00,vW00,Wittman00}.

These first detections were based on small data sets, and consequently
the accuracy of the measurements was low. However, the amount of data
is increasing rapidly \citep[e.g.,][]{Bacon02,Maoli01,Hoekstra02a,
Hoekstra02b,Refregier02,vW01,vW02}. Current data sets cover a few to
tens of degrees on the sky. Planned surveys, such as the CFHT Legacy
Survey, which will obtain deep imaging data for 170 deg$^2$ in five
filters, will be another major leap forward. The constraints on
cosmological parameters are tightening rapidly, and the consistency of
the most recent results \citep{Bacon02,Hoekstra02b,Refregier02,vW02}
suggest that cosmic shear can play an important role in this era of
``precision cosmology''.

Another important application of weak lensing is the study of the
distribution of the dark matter with respect to the galaxies.  For
instance, one can study the statistical properties of the galaxy and
dark matter distributions: the biasing relations
\citep{Hoekstra01b,Hoekstra02c}. Such measurements can provide
important constraints on models of galaxy formation. Weak lensing can
also be used to study the properties of the dark matter halos
surrounding galaxies. Rotation curves of spiral galaxies have provided
important evidence for the existence of dark matter halos
\citep[e.g.,][]{AS86}. Also, strong lensing studies of multiple imaged
systems require massive halos to explain the oberved image
separations. However, both methods provide mainly constraints on the
halo properties at relatively small radii.

The weak lensing signal can be measured out to large projected
distances, and in principle it can be a powerful probe of the
potential at large radii, constraining the amount of dark matter
\citep[e.g.,][]{Brainerd96,Griffiths96,Ian96,Hudson98,Fischer00,
thesis,Wilson01,McKay01}, the extent of the dark matter halos
\citep[e.g.,][]{Brainerd96,Hudson98,Fischer00,thesis,Hoekstra02d,Hoekstra02e}
or the average shape of the halos \citep[][]{Hoekstra02e}.

So far, we have assumed that we can apply the theory of General
Relativity to describe gravity. Consequently, the observational data
require the existence of dark matter.  However, at the large scales
probed in this paper, alternative theories have been proposed to
explain the shapes of galaxy rotation curves without the use of dark
matter. In particular Modified Newtonian Dynamics (MOND; \citet{M83})
succesfully reproduces rotation curves using only visible
matter. Ideally, one would like to test such alternative theories
far away from the visible matter. Potentially, weak lensing can place
tight constraints on such theories \citep{MT01a,MT01b}.

In this review, we will address various applications of, and
recent developments in weak gravitational lensing. In Section~2 we
concentrate on the cosmic shear measurements, and Section~3 deals with
galaxy biasing. The constraints on the halos of galaxies from weak
lensing are described in Section~4. In Section~5 we discuss how the
lensing data can be used to place limits on alternative theories of
gravity.

Throughout the paper we will make use of results based on $R_C$-band
imaging data from the Red-Sequence Cluster Survey
\citep[RCS;][]{Gladders00, Yee01}.  We refer the reader to
\citet{Hoekstra02a} for a thorough discussion of the weak lensing
analysis of these data.

\section{Lensing by large scale structure}

The measurement of the coherent distortions of the images of
faint galaxies caused by weak lensing by intervening large
scale structures provides a direct way to study the 
statistical properties of the growth of structure in the universe.
Compared to many other methods, weak lensing probes the matter
directly, regardless of the light distribution. In addition
it provides measurements on scales from the quasi-linear
to the non-linear regime where comparisons between observations
and predictions are still limited.

We provide a basic description of the theory of weak lensing by large
scale structure. Detailed discussions on this subject can be found
elsewhere \citep[e.g.,][]{Sea98,BS01}. Current measurements of cosmic
shear have concentrated on the two-point statistics, which can be
related to the convergence power spectrum, which is defined as

\begin{equation} 
P_\kappa(l)=\frac{9 H_0^4 \Omega_m^2}{4 c^4}
\int\limits_0^{w_H}dw \left(\frac{\bar W(w)}{a(w)}\right)^2
P_\delta\left(\frac{l}{f_K(w)};w\right),
\end{equation}

where $w$ is the radial (comoving) coordinate, $w_H$
corresponds to the horizon, $a(w)$ the cosmic scale factor, and
$f_K(w)$ the comoving angular diameter distance. As shown by
\citet{Jain97} and \citet{Sea98} it is necessary to use the non-linear
power spectrum in equation~(1). This power spectrum is derived from
the linear power spectrum following the prescriptions from
\citet{PD96}. We note that this approach might not be accurate enough
for future measurements \citep[e.g.,][]{vW01b,vW02}.

$\bar W(w)$ is the source-averaged ratio of angular diameter distances
$D_{ls}/D_{s}$ for a redshift distribution of sources $p_b(w)$:

\begin{equation}
\bar W(w)=\int_w^{w_H} dw' p_b(w')\frac{f_K(w'-w)}{f_K(w')}.
\end{equation}

Hence, it is important to know the redshift distribution of the
sources, in order to relate the observed lensing signal to
$P_\kappa(l)$. 

Until recently, the top-hat variance was the most widely used
two-point statistic. It is related to the convergence power spectrum
through

\begin{equation}
\langle\gamma^2\rangle(\theta)=2\pi\int_0^\infty
dl~l~P_\kappa(l)\left[\frac{J_1(l\theta)}{\pi l \theta}\right]^2,
\end{equation}

\noindent where $\theta$ is the radius of the aperture used to compute
the variance, and $J_1$ is the first Bessel function of the first
kind. 

Another useful statistic is the aperture mass, which is defined
as \citep[e.g.,][]{Kaiser94,Sea98}

\begin{equation}
M_{\rm ap}(\theta)=\int d^2\phi U(\phi) \kappa(\phi).
\end{equation}

Provided $U(\phi)$ is a compensated filter, i.e., $\int d\phi \phi
U(\phi)=0$, with $U(\phi)=0$ for $\phi>\theta$, the aperture mass can
be expressed in term of the observable tangential shear $\gamma_{\rm
t}$ using a different filter function $Q(\phi)$ (which is a function
of $U(\phi)$). Using the choice of filter functions from \citet{Sea98},
the variance of the aperture mass $\langle M_{\rm ap}^2\rangle$ is
related to the power spectrum through

\begin{equation}
\langle M_{\rm ap}^2\rangle=2\pi\int_0^\infty dl~l
P_\kappa(l)\left[\frac{12}{\pi (l \theta)^2} J_4(l \theta)\right]^2,
\end{equation}

\noindent where $J_4$ is the fourth-order Bessel function of the first
kind. The measurement of $\langle M_{ap}^2\rangle$ is of particular
interest, because the measurements at different scales are only
weakly correlated. 

The first detections of cosmic shear
\citep{Bacon00,Kaiser00,Maoli01,vW00, Wittman00} actually computed the
excess variance in apertures directly from the data. Some more recent
studies have used the same technique
\citep{Bacon02,Hoekstra02a,Refregier02,vW01}. 

However, recent studies show that an optimal use of the data is to
measure the shear correlation functions from the data \citep{Pen02,
vW02,Hoekstra02b}. These correlation functions can be related to the
various two-point statistics \citep{Crittenden02,Pen02}. In addition,
this approach allows one to split the signal into two components: an
``E''-mode, which is curl-free, and a ``B''-mode, which is sensitive
to the curl of the shear field.  Gravitational lensing arises from
a gravitational potential, and hence it is expected to produce a
curl-free shear field. Thus, the ``B''-mode can be used to quantify
the level of systematics involved in the measurement.  The
decomposition is naturally carried out by using the aperture mass
statistic $M_{\rm ap}$.

Several sources of B-mode have been identified.  For instance, simple
models describing the intrinsic alignments of galaxies predict a small
B-mode \citep[e.g.,][]{Crittenden02}, although the amplitude is still
uncertain.  Hence, any measured B-mode is dominated by residual
systematics in the data (e.g., imperfect correction of the PSF
anisotropy) or intrinsic alignments. The effect of intrinsic aligments
can be minimized by selecting galaxies with a broad redshift
distribution. In future surveys, with photometric redshift information
for the galaxies, the contribution from intrinsic aligments can be
removed completely by correlating the shapes of galaxies with
different redshifts.

\begin{figure*}
\leavevmode
\centering
\hbox{
\epsfxsize=\hsize
\epsffile[18 315 558 710]{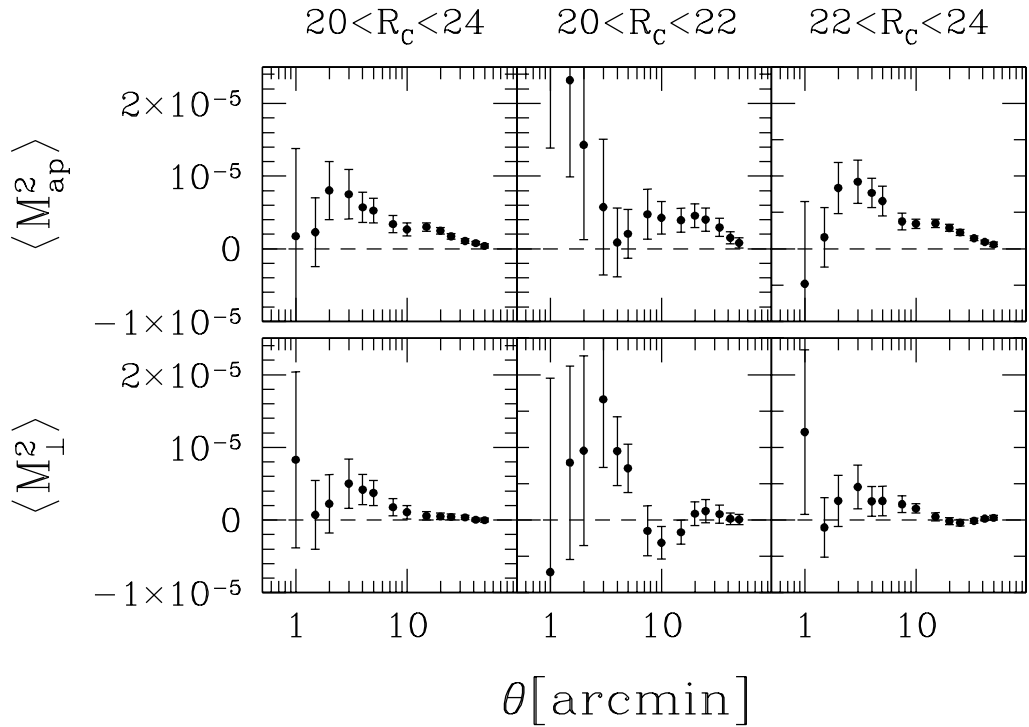}} 
\vspace{-0.3cm}
\caption{
\footnotesize The upper panels show the measured variance
of the aperture mass $\langle M_{\rm ap}^2\rangle$ as a function of
aperture size $\theta$ for different samples of source galaxies.  This
signal corresponds to the E-mode. The lower panels show the variance
$\langle M_\perp^2\rangle$ when the phase of the shear is increased by
$\pi/2$, and corresponds to the B-mode. The error bars indicate the
$1\sigma$ statistical uncertainty in the measurements, and have been
derived from the field-to-field variation of the 13 RCS patches (thus the
error bars include cosmic variance). Note that the points are slightly
correlated. We detect a significant B-mode on scales $5-10$
arcminutes. On scales larger than 10 arcminutes the B-mode
vanishes. The sample of bright galaxies $(20<R_C<22)$ should not be
affected significantly by systematics because their sizes are large
compared to the PSF. Therefore the significant B-mode at scales of a
few arcminutes is likely to be caused by intrinsic alingments.  To
minimize the effect of intrinsic alignments on our cosmological
parameter estimation, we will use the sample of galaxies with
$22<R_C<24$ to this end.
\label{mapall}}
\end{figure*}

Figure~\ref{mapall} shows the results of such a decomposition into
``E'' and ``B''-modes. The analysis, which is described in detail in
\citet{Hoekstra02b}, uses 53 deg$^2$ of RCS imaging data.  For all
three samples we find that the ``B''-mode (lower panels) vanishes on
scales larger than $\sim 10$ arcminutes, suggesting that neither
observational distortions or intrinsic alignments of sources have
corrupted our measurements.

On smaller scales we detect a significant ``B''-mode. Interestingly,
we detect a significant signal for the bright galaxies
$(20<R_C<22)$. These galaxies have sizes that are large compared to
the PSF, and therefore they are less affected by residual
systematics. Intrinsic alignments are expected to be particularly
important for these bright galaxies, and we therefore conclude that
(at least part of) the observed ``B''-mode is likely to be caused by
intrinsic alignments.

\begin{figure}
\leavevmode
\centering
\hbox{
\epsfysize=6.5cm
\epsfbox{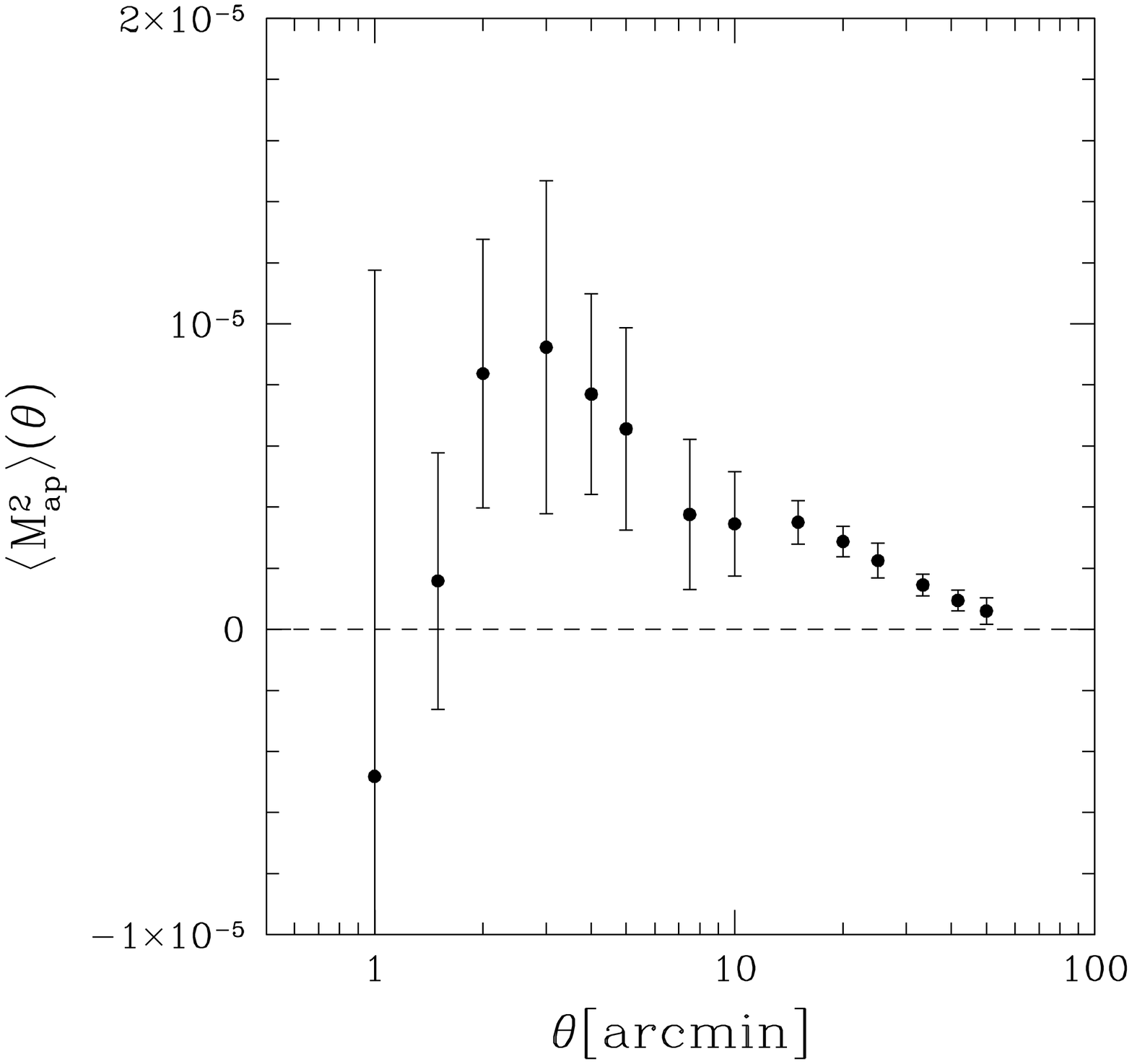}
\epsfysize=6.5cm
\epsfbox{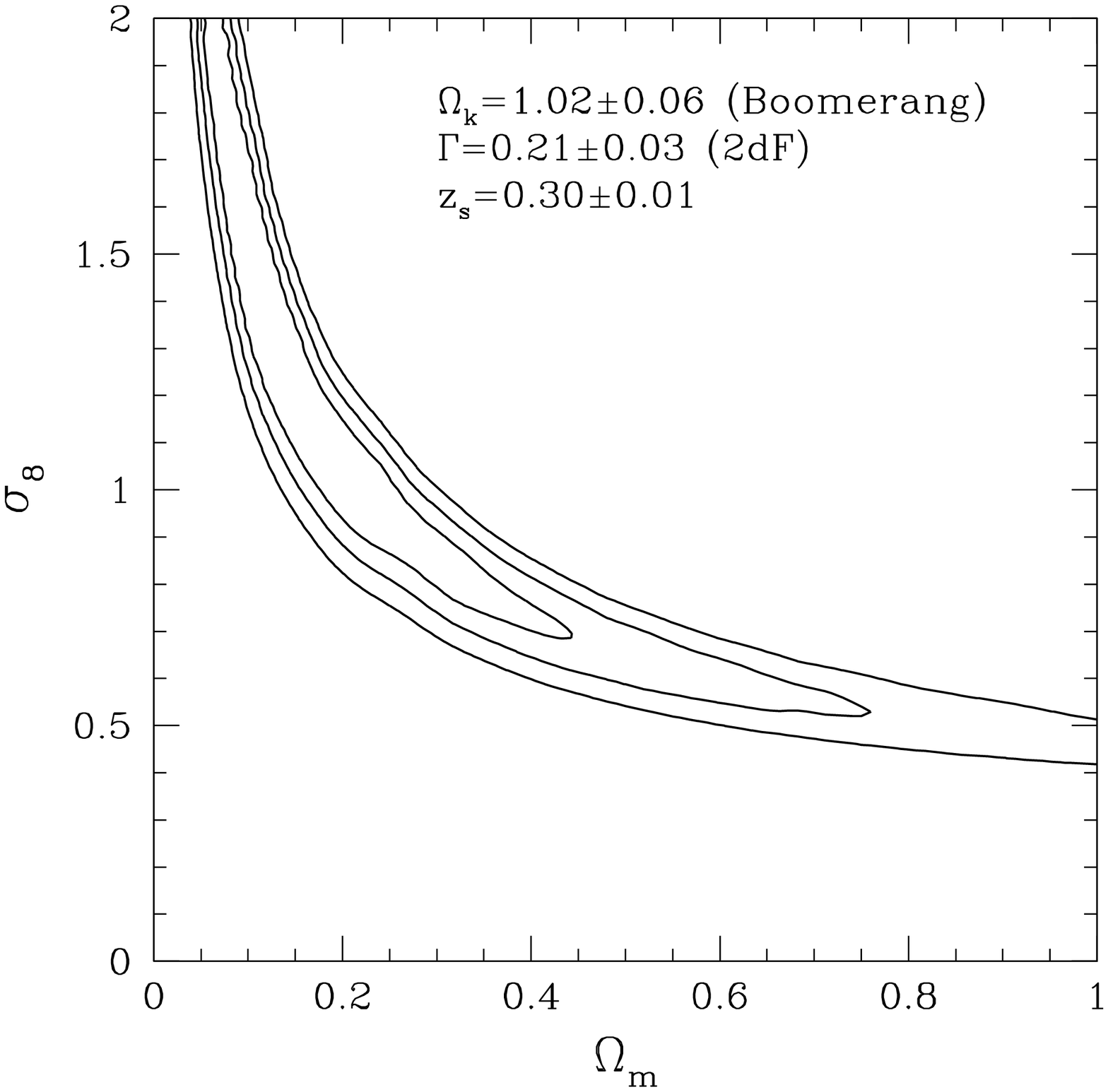}}
\vspace{-0.3cm}
\caption{\footnotesize {\it Left panel:} The measured variance
of the aperture mass $\langle M_{\rm ap}^2\rangle$ as a function of
aperture size $\theta$ for source galaxies with $22<R_C<24$. The
error bars have been increased to account for the unknown correction
for the ``B''-mode observed in Figure~\ref{mapall}. {\it Right panel:}
Contraints on $\Omega_m$ and $\sigma_8$ using priors Gaussian priors
on $\Gamma$, $\Omega_{\rm tot}=\Omega_m+\Omega_\Lambda$, and the source
redshift distribution $z_s$ (which gives $\langle
z\rangle=0.59\pm0.02$).  The likelihood contours have been derived by
comparing the measurements to CDM models with $n=1$.  For $\Gamma$ we
used the constraints from the 2dF survey $\Gamma=0.21\pm0.03$ 
\citep{Peacock01,Efstathiou01}, and for $\Omega_{\rm tot}$ we used the
Boomerang constraints $\Omega_{\rm tot}=1.02\pm0.06$ \citep{Netterfield01}. 
The contours indicate the 68.3\%, 95.4\%, and 99.7\%
confidence limits on two paramaters jointly.
\label{mapfaint}}
\end{figure}

The next step is to relate the observed signal to estimates of
cosmological parameters. Our current understanding of the redshift
distribution of the source galaxies is sufficient to obtain accurate
constraints. The most accurate results are derived when one uses
external priors on the parameters. Useful constraints come from
redshift surveys and the temperature fluctuations in the CMB.  Current
weak lensing measurements provide joint constraints on $\Omega_m$ and
$\sigma_8$, and Figure~\ref{mapfaint} shows the results from
\citet{Hoekstra02b}. However, measurements of higher order statistics,
such as the skewness, can be used to break the degeneracies between
$\Omega_m$, and $\sigma_8$ \citep[e.g.,][]{vW99}. Recently the first
detection of a measure of skewness was reported
\citep{Bernardeau02a,Bernardeau02b}, and the prospects for
measurements of higher order statistics are good.

In Table~\ref{tab_comp} we list the estimates of $\sigma_8$ (for
$\Omega_m=0.3$ and $\Gamma=0.21$) as derived by the most recent weak
lensing surveys \citep{Bacon02,Hoekstra02b,Refregier02,vW02}. Despite
the fact that these results have been obtained using a wide variety of
telescopes (both ground based and HST), filters, and integration
times, the agreement is remarkable.

We note that the value of $\sigma_8$ determined from the RCS data
is essentially constrained by the measurements at scales larger
than 10 arcminutes, where the ``B''-mode is negligible. The result
from Van Waerbeke et al. (2002) might be biased high, because
of their large scale ``B''-mode. Both \cite{Bacon02} and \cite{Refregier02}
do not separate their signal into ``E'' and ``B'-modes,
and therefore it might also include some residual systematics. 

A widely used method to determine the normalization of the power
spectrum uses the number density of rich clusters of galaxies
\citep[e.g.,][]{Borgani01,Carlberg97,Eke96,Fan98,Pen98a,Pierpaoli01,
Reiprich01,Seljak01,Viana01}. Such systems are rare, and as a result a
very sensitive probe of $\sigma_8$, provided one can determine their
mass. The derived values for $\sigma_8$ from this technique range from
values as low as $0.61\pm0.05$ \citep{Viana01} to values around unity
\citep[e.g.,][]{Fan98,Pen98a,Pierpaoli01}. The statistical error bars
on most of these measurements are small, and hence the large spread in
values suggests an underlying large systematic uncertainty.

\begin{table}
\begin{center}
\begin{tabular}{ll}
\hline
\hline
Cosmic shear survey & $\sigma_8$ \\
\hline
\citet{Hoekstra02b} (RCS)	& $0.86^{+0.04}_{-0.05}$ \\
\citet{Bacon02}			& $0.97^{+0.10}_{-0.09}$ \\
\citet{Refregier02}		& $0.94\pm 0.14$\\
\citet{vW02}			& $0.98\pm 0.06$ \\
\hline
\hline
\end{tabular}
\caption{\footnotesize Values of $\sigma_8$ and 68\% confidence
intervals as derived from 4 independent cosmic shear measurements (adopting
$\Omega_m=0.3$, $\Omega_\Lambda=0.7$, and $\Gamma=0.21$.
\label{tab_comp}}
\end{center}
\end{table}

\section{Galaxy biasing}

The growth of structures in the universe via gravitational instability
is an important ingredient in our understanding of galaxy formation.
However, the connection to observations is not straightforward, as we
need to understand the relation between the dark matter distribution
and the galaxies themselves. Galaxy formation is a complex process,
and it is not guaranteed a priori that this relation, referred to as
galaxy biasing, is a simple one. The bias might be non-linear, scale
dependent or stochastic. In the simplest case, linear, deterministic
biasing, the relation between the dark matter and the galaxies can be
characterized by a single number $b$ \citep[e.g.,][]{Kaiser87}.

Most observational constraints of biasing come from dynamical studies
\citep[see ][]{Strauss95} which probe relatively large scales ($10
h_{50}^{-1}$~Mpc or more). Recent estimates on these scales suggest
values of $b\sim 1$ for $L_*$ galaxies \citep[e.g.,][]{Peacock01,Verde01}.
On smaller scales some constraints come from measurements of the
galaxy two-point correlation function, which is compared to the (dark)
matter correlation function computed from numerical simulations. 
These studies indicate that the bias parameter $b\simeq 0.7$ on scales
less than $\sim 2 h_{50}^{-1}$ Mpc
$(\Omega_m=0.3,~\Omega_\Lambda=0.7)$. On larger scales $b$ increases
to a value close to unity \citep{Jenkins98}
Although this procedure provides useful information about the bias
parameter $b$ as a function of scale, it does rely on the assumptions
made for the numerical simulations. In addition, it cannot be used to
examine how tight the correlation between the matter and light
distribution is. To do so, we need to measure the galaxy-mass
cross-correlation coefficient $r$, which is a measure of the amount of
stochastic and non-linear biasing \citep[e.g.,][]{Pen98b,Dekel99,Somerville01}.

Redshift surveys can be used to determine the relative values of $b$
and $r$ for different galaxy types \citep{Tegmark99,Blanton00}, but
weak gravitational lensing provides the only direct way to measure the
galaxy-mass cross-correlation function
\citep[e.g.,][]{Fischer00,Wilson01,McKay01, Hoekstra01b}.
\cite{Fischer00} used the SDSS commissioning data to measure the
galaxy-mass correlation function and their results suggested an
average value of $b/r\sim 1$ on submegaparsec scales. This approach
has been explored by \cite{Guzik01} who used semi-analytic models of
galaxy formation combined with N-body simulations. Their results
suggest that the cross-correlation coefficient is close to unity.

To study the galaxy biasing we use a combination of the galaxy and
mass auto-correlation functions, as well as the cross-correlation
function.  A common definition of the ``bias'' parameter is the ratio
of the variances of the galaxy and dark matter densities, which is the
definition we will use here. In the case of deterministic, linear
biasing the galaxy density contrast $\delta_g$ is simply related to
the mass density contrast $\delta$ as $\delta_g=b\delta$
\citep{Kaiser87}, and the ratio of the variances is the only relevant
parameter.  However, the bias relation is likely to be more
complicated as it depends on the process of galaxy formation, and
might be stochastic, non-linear or both. We allow for non-linear
stochastic biasing by including the galaxy-mass cross-correlation
coefficient $r$ \citep[e.g.,][]{Pen98b,Dekel99,Somerville01}. The
cross-correlation coefficient $r$ mixes non-linear and stochastic
effects \citep{Dekel99,Somerville01} which we currently cannot
disentangle with weak lensing.

The bias parameters are related to the observed correlation
functions through \citep{Hoekstra02c}

\begin{equation}
b^2=f_1(\theta_{\rm ap},\Omega_m,\Omega_\Lambda)\times
\frac{\langle{N}^2(\theta_{\rm ap})\rangle}
{\langle M_{\rm ap}^2(\theta_{\rm ap})\rangle},
\end{equation}

\noindent and

\begin{equation}
r=f_2(\theta_{\rm ap},\Omega_m,\Omega_\Lambda)\times
\frac{\langle M_{\rm ap}(\theta_{\rm ap}){N}(\theta_{\rm ap})\rangle}
{\sqrt{\langle {N}^2(\theta_{\rm ap})\rangle
\langle M_{\rm ap}^2(\theta_{\rm ap})\rangle}},
\end{equation}

\noindent where $f_1$ and $f_2$ depend on the assumed cosmological
model and the redshift distributions of the lenses and the sources
\citep[see][]{vW98,Hoekstra02c}. The values of $f_1$ and $f_2$,
however, depend minimally on the assumed power spectrum and the
angular scale.

There is no reason for $b$ or $r$ to be constant with scale, but as
long as $b$ and $r$ vary slowly with scale, we can still infer the
bias parameters from Eqs.~(3) and (4). The $M_{\rm ap}$ statistic is
sensitive to a fairly small range in $k$ in Fourier space for a given
aperture size. If the galaxy power spectrum $P_{\rm gg}(k)$ can be
related to the matter power spectrum $P_{\rm mm}(k)$ through $P_{\rm
gg}(k)=b^2(k)P_{\rm mm}(k)$, we essentially measure the average value
of $b(k)$ in the $k$-range probed by the $M_{\rm ap}$ statistic
\citep{vW98,Hoekstra02c}. In other words, $M_{\rm ap}$ is a pass-band
filter, which also explains why different scales are only slightly
correlated \citep{Sea98}.

\begin{figure}
\leavevmode
\centering
\hbox{
\epsfysize=13cm
\epsfbox{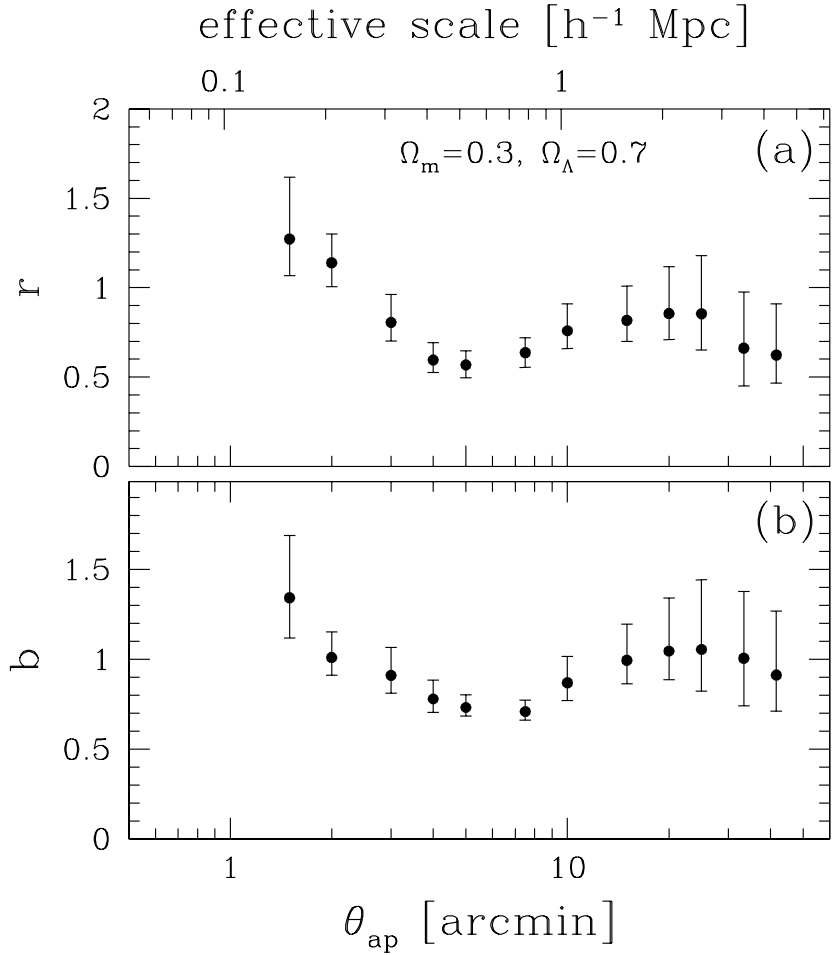}}
\vspace{-0.3cm}
\caption{\small (a) The measured value of the galaxy-mass
cross correlation coefficient $r$ as a function of scale for the
$\Lambda$CDM cosmology. (b) The bias parameter $b$ as a function
of scale. The upper axis indicates the effective physical scale probed by the
compensated filter at the median redshift of the lenses
$(z=0.35)$. The error bars correspond to the 68\% confidence intervals.
Note that the measurements at different scales are slightly
correlated.
\label{bias}}
\end{figure}

\cite{Hoekstra02c} studied the bias properties of galaxies with
$19.5<R_C<21$. They combined the weak lensing measurements from the
RCS and the VIRMOS-DESCART survey \citep[e.g.,][]{vW02} to obtain the
first direct measurements of the bias parameter $b$ and the
galaxy-mass cross-correlation coefficient $r$ as a function of scale
(with scales ranging from 0.1 to 4 $h^{-1}$ Mpc).  The results of this
analysis are presented in Figure~\ref{bias}.  \cite{Hoekstra02c} find
that both $b$ and $r$ vary with scale for this sample of galaxies. The
low value of $r$ on scales $\sim 0.5-1 h^{-1}$ Mpc suggests
significant stochastic biasing and/or non-linear biasing.
Although both $b$ and $r$ vary with scale, the ratio $b/r$ turns
out to be almost constant with scale. \cite{Hoekstra02c} find
that $b/r=1.090\pm0.035$ out to $7h^{-1}$ Mpc. 

Models of galaxy formation predict the relation between the galaxies
and the underlying dark matter distribution. The two commonly used
approaches are hydrodynamic simulations
\citep[e.g.,][]{Blantonea00,Yoshikawa01} or
\citep[e.g.,][]{Kauffmann99a, Kauffmann99b,Somerville01,Guzik01}. A
comparison of the models with the lensing results can provide accurate
constraints. Unfortunately, current models of galaxy formation are
limited to relatively bright (i.e., massive) galaxies, because of the
mass resolution of the numerical simulations. Most of the galaxies in the
sample studied by \cite{Hoekstra02c} are too faint to be included in
the simulations. This severely limits a direct comparison. However,
with available multi-color data the weak lensing results can be
matched better to the models, and vice versa, future models should
include lower mass galaxies.

\section{Constraints on the halos of galaxies}

Rotation curves of spiral galaxies \citep[e.g.,][]{AS86} and strong
lensing analyses of multiple imaged sources have provided strong
evidence for the existence of massive dark matter halos around
galaxies.  However, the lack of visible tracers hampers our knowledge
of the gravitational potential at large projected distances.  Only
satellite galaxies \citep[e.g.,][]{ZW94} provide a way to
probe the outskirts of isolated galaxy halos.

However, the weak gravitational lensing signal induced by such
galaxies can be measured out to large projected distances, thus
providing a powerful probe of the potential at large radii, which can
be used to study the extent of the dark halos \citep[e.g.,][]{Brainerd96,
Hudson98, Fischer00, thesis, Hoekstra02d, Hoekstra02e}.

The lensing signal induced by an individual galaxy is too low to be
detected, and one has to study the ensemble averaged signal around a
large number of lenses. Early measurements of the galaxy-galaxy
lensing signal were limited by the small number of lenses/sources
\citep[e.g.,][]{Brainerd96, Hudson98}, because of the lack of panoramic
cameras. Recently the accuracy with which the lensing signal can be
measured has improved significantly. \citet{Fischer00} measured a very
significant signal, and \citet{Wilson01} studied the signal around early
type galaxies as a function of redshift. \citet{McKay01} used the
redshift information from the SDSS to study the galaxy-galaxy lensing
signal as a function of galaxy properties.

\begin{figure}
\begin{center}
\leavevmode 
\hbox{%
\epsfxsize=8cm
\epsffile[20 175 570 690]{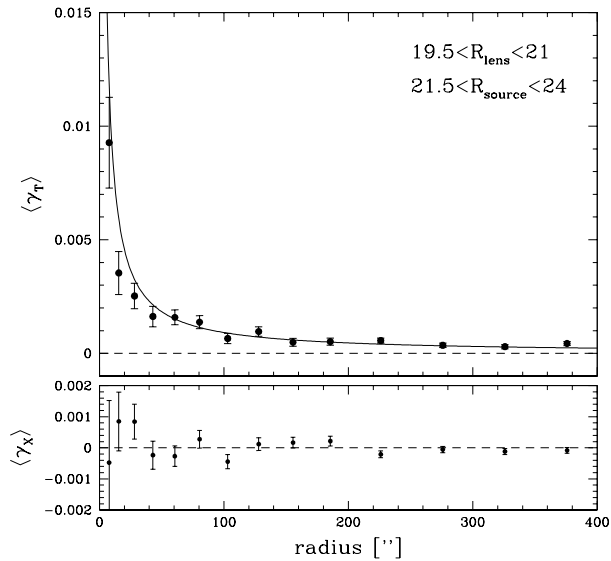}} 
\caption{\footnotesize (a) The ensemble averaged
tangential distortion as a function of radius around the sample of
lens galaxies. The measurements have been corrected for the presence
of source galaxies associated with the lens galaxies. (b) The signal
when the phase of the distortion is increased by $\pi/2$: no signal
should be present if the signal in (a) is due to lensing. The vertical
error bars indicate the $1\sigma$ errors\label{gtprof}}
\end{center}
\end{figure}

Here we present some preliminary results based on 45 deg$^2$ of RCS
$R_C$ band data. The detailed analysis is described in
\citet{Hoekstra02e}, and a description of the method can be found in
\citet{thesis}. We use galaxies with $19.5<R_C<21$ as lenses, and
galaxies with $21.5<R_C<24$ as sources which are used to measure the
lensing signal. The simplest approach to galaxy-galaxy lensing is to
measure the ensemble averaged tangential distortion as a function of
radius around the sample of lenses. The result for our data is
presented in Figure~\ref{gtprof}a. The signal when the phase of the
shear is increased by $\pi/2$ is presented in
Figure~\ref{gtprof}b. This signal is consistent with zero, indicating
that observational systematics have been successfully corrected for.

We fit a SIS model to the tangential distortion which yields $\langle
r_E\rangle=0\farcs{129}\pm0\farcs{011}$. Given the redshifts of the
lenses and the sources this corresponds to a velocity dispersion of
$\langle\sigma^2\rangle^{1/2}=123\pm5$ km/s. The corresponding
circular velocity can be obtained using $V_c=\sqrt{2}\sigma$. 

The tangential shear profile is a convolution of the actual halo
profile around galaxies and the clustering properties of the
lenses. \citet{Fischer00} used this tangential shear profile, taking into
account the clustering of the lenses, to obtain constraints on the
extent of the dark matter halos surrounding the lenses. They conclude
that the halos are large, but they were unable to obtain tight
constraints. In Section~4.1 we show the results of a maximum
likelihood analysis of the data, and demonstrate that this approach
gives much better constraints on the sizes of halos.

An application of galaxy-galaxy lensing that is still in its 
early stage is the measurement of the shapes of galaxy halos.
If the halos are flattened, the lensing signal will be
slightly anisotropic. Provided the halos and the galaxies
are aligned, we can measure this angular dependence
of the tangential shear. In Section~4.2, we present the first
results of such an analysis.

\subsection{Sizes of galaxy halos}

For an isolated lens, the induced lensing signal is purely tangential.
This is no longer the case for an ensemble of lenses (which also
cluster): the other lenses cause small perturbations, and the shear is
no longer tangential.  To study the sizes of halos around galaxies
\citet{Fischer00} used only the tangential component. However,
\citet{thesis} was able to obtain better constraints using a maximum
likelihood analysis on their imaging data of CNOC2 fields.  The
maximum likelihood analysis uses a parameterized model for the mass
distribution of individual galaxies is compared to the data.  This has
the advantage that one uses the information contained in both
components of the distortion. It implicitely assumes that all
clustered dark matter is associated with galaxies.  Furthermore, the
contributions from faint neighboring galaxies are not included in the
model, and as such, the results reflect the average density profile
around the lenses.

A useful model to describe a truncated halo is \citep{SR97, thesis}

\begin{equation}
\Sigma(r)=\frac{\sigma^2}{2Gr}\left(1-\frac{r}{\sqrt{r^2+s^2}}\right),
\end{equation}

where $s$ is a measure of the truncation radius (i.e., the scale
where the profile steepens). The total mass of this model is finite,
and half of the mass is contained with $r=\frac{3}{4}s$.

As before we assume that $\sigma\propto L_B^{1/4}$. We also have to
assume a scaling relation for the truncation parameter $s$. Currently,
no such observational constraints exist, although galaxy-galaxy
lensing analyses with redshifts for the lenses can be used to this
end. As shown by \citet{thesis} different scaling relations lead to
somewhat different values for $s$. Also one can compare the
predictions from N-body simulations \citep{NFW} to the data.  The aim here
is not to derive the definitive value of $s$, but to demonstrate that
weak lensing can constrain the ``extent'' of the dark matter halos.

\begin{figure}
\begin{center}
\leavevmode 
\hbox{%
\epsfxsize=8cm
\epsffile[20 175 570 690]{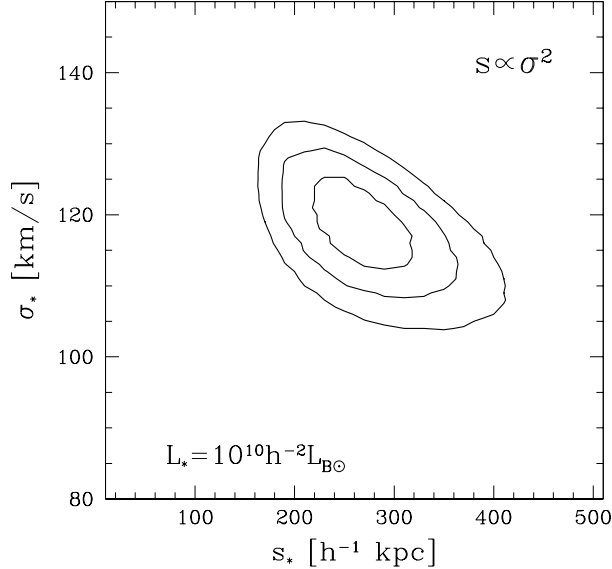}} 
\caption{\footnotesize Likelihood contours for the velocity dispersion
$\sigma_*$ and truncation parameter $s_*$ of a galaxy with a fiducial
luminosity $L_*=10^{10}h^{-2}{\rm L}_{{\rm B}\odot}$. To derive this
result we assumed that the velocity dispersion $\sigma$ scales $\propto
L_{\rm B}^{1/4}$, and that the truncation parameter $s$ scales
$\propto L_{\rm B}^{1/2}$ (or $\propto \sigma^2$).  The contours
indicate the 68.3\%, 95.4\%, and the 99.7\% confidence on two
parameters jointly.
\label{chi2}}
\end{center}
\end{figure}

Currently, we do not have (photometric) redshift information
for the lens galaxies in the RCS fields. To derive constraints
on the dark matter halos we follow \citet{thesis}, and make
mock catalogs of lenses where the redshifts are drawn from
the CNOC2 redshift survey \citep{Yee00} on the basis of their
$R_C$-band magnitude. This provides a very crude redshift estimate.
We found that the derived results did not change much between
realisations.

In the analysis we assumed that the velocity dispersion $\sigma$
scales $\propto L_{\rm B}^{1/4}$, and that the truncation parameter
$s$ scales $\propto L_{\rm B}^{1/2}$ (or $\propto
\sigma^2$). Figure~\ref{chi2} shows the results of maximum likelihood
analysis for a galaxy with a fiducial luminosity of
$L_*=10^{10}h^{-2}{\rm L}_{{\rm B}\odot}$.  For such a galaxy we find
$\sigma_*=118^{+5}_{-4}$, in fair agreement with other estimates. In
addition, the average extent of the dark matter halo is well
constrained, and we obtain $s_*=265^{+33}_{-25} h^{-1}$ kpc, and a
99.7\% confidence upper limit of $s_*<390 h^{-1}$ kpc.

\subsection{Shapes of galaxy halos}

Numerical simulations of cold dark matter result in triaxial halos,
with typical ellipticity of $\sim 0.3$ \citep[e.g.,][]{Dubinski91}. In
the context of collisionless cold dark matter, the theoretical
evidence for flattened halos is quite strong. However, the
observational evidence is still limited. \citet{Sackett99} gives an
overview of current constraints which probe the vertical potential on
scale $\le$ 15 kpc. These results suggest an average value of
$c/a=0.5\pm0.2$ (where $c/a$ is the ratio of the shortest to longest
principle axis of the halo).

Current constraints are mainly limited by the lack of visible tracers
that can probe the gravitational potential around galaxies.  Hence,
galaxy-galaxy lensing is potentially the most powerful way to derive
constraints on the average shape of dark matter halos. As such, it
provides an important test of the collisionless cold dark matter
paradigm.  \citet{Brainerd00} and \cite{Natarajan00} examined the data
requirements needed for a detection of the halo flattening.  Although
their estimates appear to be too optimistic, it is clear that large
amounts of data are required.  The measurement of the average shape of
the dark matter halos is much more difficult than the previous
measurements: the galaxy-galaxy lensing signal is small, and now one
needs to measure an even smaller azimuthal variation. Furthermore one
has to assume that the galaxy and its halo are aligned, which appears
to be a reasonable assumption.

\begin{figure}
\begin{center}
\leavevmode 
\hbox{%
\epsfxsize=8cm
\epsffile{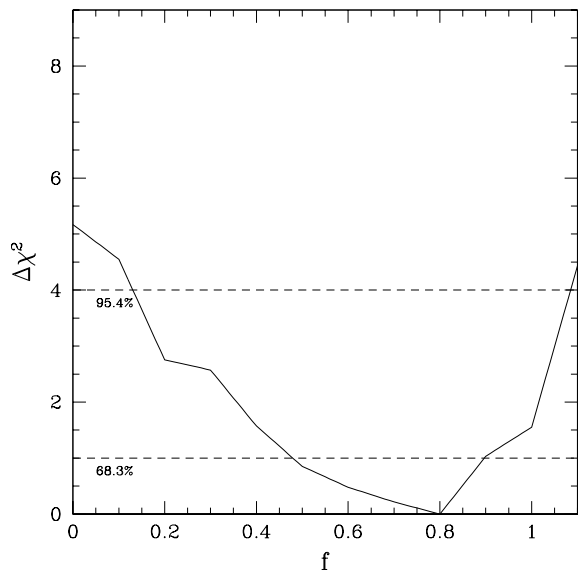}} 
\caption{\footnotesize $\Delta \chi^2$ as a function of $f$. We have
assumed that the ellipticity of the halos is related to the observed
ellipticity of the lens as $e_{\rm halo}=f e_{\rm lens}$.  We have
indicated the 68.3\% and 95.4\% confidence intervals.  Round halos
$(f=0)$ are excluded with 99\% confidence.  The average ellipticity of
the galaxies is $\langle e_{\rm lens}\rangle=0.261$ and hence the
average halo ellipticity is $\langle e_{\rm
halo}\rangle=0.21^{+0.02}_{-0.08}$ (68.3\% confidence), which
corresponds to an average projected axis ratio of
$c/a=0.65^{+0.12}_{-0.02}$ (68.3\% confidence).
\label{flat_halo}}
\end{center}
\end{figure}

Our approach is different from that of \citet{Brainerd00} and
\cite{Natarajan00}, who propose to study the azimuthal variation in
the tangential shear around the lenses. Instead we use a maximum
likelihood analysis to derive constraints on the shapes of dark matter
halos. We note that this is a first attempt to do so, and more
detailed work is required. We simply assume that the (projected)
ellipticity of the dark matter halo is related to the shape of the
galaxy as $e_{\rm halo}=f e_{\rm lens}$, and use the best fit
parameters for the velocity dispersion and truncation. The
ellipticities of the lenses have been corrected for the PSF, similar
to the sources. The lensing signal is dominated by the contribution
from early type galaxies, and therefore our parameterisation of the
halo shape is reasonable.

The resulting $\Delta\chi^2$ as a function of $f$ is presented in
Figure~\ref{flat_halo}. The main conclusion from this analysis is that
spherical halos are excluded with 99\% confidence. Although the
inferred ellipticity of dark matter halos depends on the adopted
dependence on the light distribution, we note that the exclusion
of a spherical halo is robust. Compared to the measurement
of the tangential shear as a function of radius, the result
presented here is more sensitive to residual systematics. A
detailed analysis, however, demonstrates that our result is
robust, and not caused by systematics.

The average ellipticity of the lens galaxies is $\langle e_{\rm
lens}\rangle=0.261$. The best fit value of $f=0.8$ then implies a
average projected halo ellipticity of $\langle e_{\rm
halo}\rangle=0.21^{+0.02}_{-0.08}$ (68.3\% confidence), which
corresponds to an projected axis ratio of $c/a=0.65^{+0.12}_{-0.02}$
(68.3\% confidence). Although the weak lensing yields a projected axis
ratio, the result is in fair agreement with the numerical simulations.
We note, however, that it is not straightforward to interpret this result:
the lensing signal comes from a range of galaxy types, although
typically the lensing signal is dominated by the early type galaxies,
for which our halo model might be appropriate.

\section{Limits on alternative theories of gravity}

In the context of General Relativity, we have to invoke large amounts
of dark matter to explain a wide variety of observational
results. However, at the large scales probed in this paper, no
definite tests of general relativity have been made, and alternative
theories of gravity have been proposed to explain the shapes of galaxy
rotation curves without the use of dark matter. In particular Modified
Newtonian Dynamics (MOND; \citet{M83}) succesfully reproduces rotation
curves using only visible matter.

Dynamical studies can provide useful constraints on such theories, but
only in the regions where there is visible matter. The best regime to
test such alternative theories is far away from the visible matter.
This requires a way to probe the gravitational potential on large
physical scales. Weak gravitational lensing is currently the
best approach to this end. As shown above, it can accurately probe the 
gravitational potential on scales where other methods fail.
In addition, in the absence of dark matter halos, galaxy-galaxy
lensing is simple, because the galaxies can be treated as point masses.
Unfortunately most alternative theories of gravity do not provide a
useful description of gravitational lensing. For instance, MOND lacks
a relativistic theory that can describe light deflection.  

To test alternative theories of gravity we can use two different
approaches. A measurement of the radial dependence of the lensing
signal gives the best accuracy, but it requires knowledge of the
deflection law. The most direct test is the detection of the the
azimuthal variation of the lensing signal around the lenses:
alternative theories predict an isotropic lensing signal, and the
measurement does not require knowledge of the deflection law!

In alternative theories of gravity, any anisotropy in the lensing
signal caused by the intrinsic shapes of the galaxies (i.e., the light
and gas distribition) decreases $\propto r^{-2}$, and hence is
negligible on the scales probed here. Therefore the observed 
anisotropy in the galaxy-galaxy lensing signal (see Section~4.2)
poses a serious problem for alternative theories of gravity.  More
detailed work is required to confirm this result, but new data sets
will provide much better constraints in the very near future, and
allow us to study the anisotropy a function of projected distance from
the galaxy.

The detection of the anisotropy of the lensing signal around galaxies
provides the best evidence for the existence of dark matter halos
around galaxies. For completeness we also discuss constraints on the
radial dependence of gravity, following the approach suggested by
\citet{MT01b}.

\subsection{Constraints on the radial dependence of light deflection}

\citet{MT01b} considered a ``zero-th'' order approach, by recalling
that the deflection angle in GR is twice the Newtonian value. They
simply assumed that a similar relation holds for MOND, and derive the
``deflection law''. We note, however, that this approach is rather
simplistic, and ad hoc, because the force law used by \citet{MT01b}
does not return to the Newtonian form on large scales (although with a
modified gravitational constant), which is required from general
considerations \citep[e.g.,][]{Sanders86,Walker94,White01}.

The aim of this section is to give an indication of the potential of
current data sets for the purpose of testing alternative theories of
gravity.  Along this line of reasoning, we first concentrate of the general
class of models considered by \citet{MT01a}.  One simply starts from
the deflection law for a point mass. In General Relativity, the
deflection law is

\begin{equation}
\alpha(\theta)= - \frac{\theta_{\rm E}^2}{\theta},
\end{equation}

\noindent where $\theta_{\rm E}$ is the Einstein radius of the lens.
The Einstein radius depends on the mass $M$ of the lens and the
angular diameter distances from observer to lens $(D_{l})$, observer
to source $(D_{s})$, and lens to source $(D_{ls})$:

\begin{equation}
\theta_{\rm E}=\sqrt{\frac{4GM}{c^2}\frac{D_s}{D_l D_{ls}}}.
\end{equation}

Hence, another complication of lensing in alternative theories
of gravity arises because it is not clear how the angular diameter
distances vary with redshift. However, this uncertainty should only
affect the amplitude of the lensing signal, and not its angular
dependence. 

\citet{MT01a} considered a generic deflection law, which is
parameterized as

\begin{equation}
\alpha(\theta)= - \frac{\theta_{\rm E}^2}{\theta}
\left(\frac{\theta_0}{\theta+\theta_0}\right)^{\xi-1}\label{deflaw}.
\end{equation}

For $\theta\ll\theta_0$, this deflection law reduces to the usual
deflection, as is the case for $\xi=1$. For large $\theta$ the
deflection angle decreases $\propto\theta^{-\xi}$. The case of
$\xi=0$, can be considered as the approximate MOND deflection angle.

It is straightforward to relate the deflection angle to
the observable shear. For this particular choice of deflection
angle we obtain \citep{MT01a}

\begin{equation}
\gamma(\theta)=\frac{\theta_{\rm E}^2}{\theta_0\theta^2}
\left(\frac{\xi+1}{2}\theta+\theta_0\right)
\left(\frac{\theta_0}{\theta+\theta_0}\right)^\xi.
\end{equation}

\citet{MT01a} compared this model to the observed ensemble
averaged tangential shear around galaxies in the SDSS
\citep{Fischer00}. They found a best fit value of $\xi=-0.1$ and
$\theta_0=3.7$ arcsecond. As a result, \citet{MT01a} concluded
the MOND ``predictions'' are consistent with the SDSS measurements.

However, as we will show now, a maximum likelihood analysis
can provide much better constraint. As before, we use the
measurements from the RCS. We now treat the lenses as point
masses with the deflection law given by equation~\ref{deflaw}.

\begin{figure}
\begin{center}
\leavevmode
\hbox{%
\epsfxsize=6.8cm
\epsffile{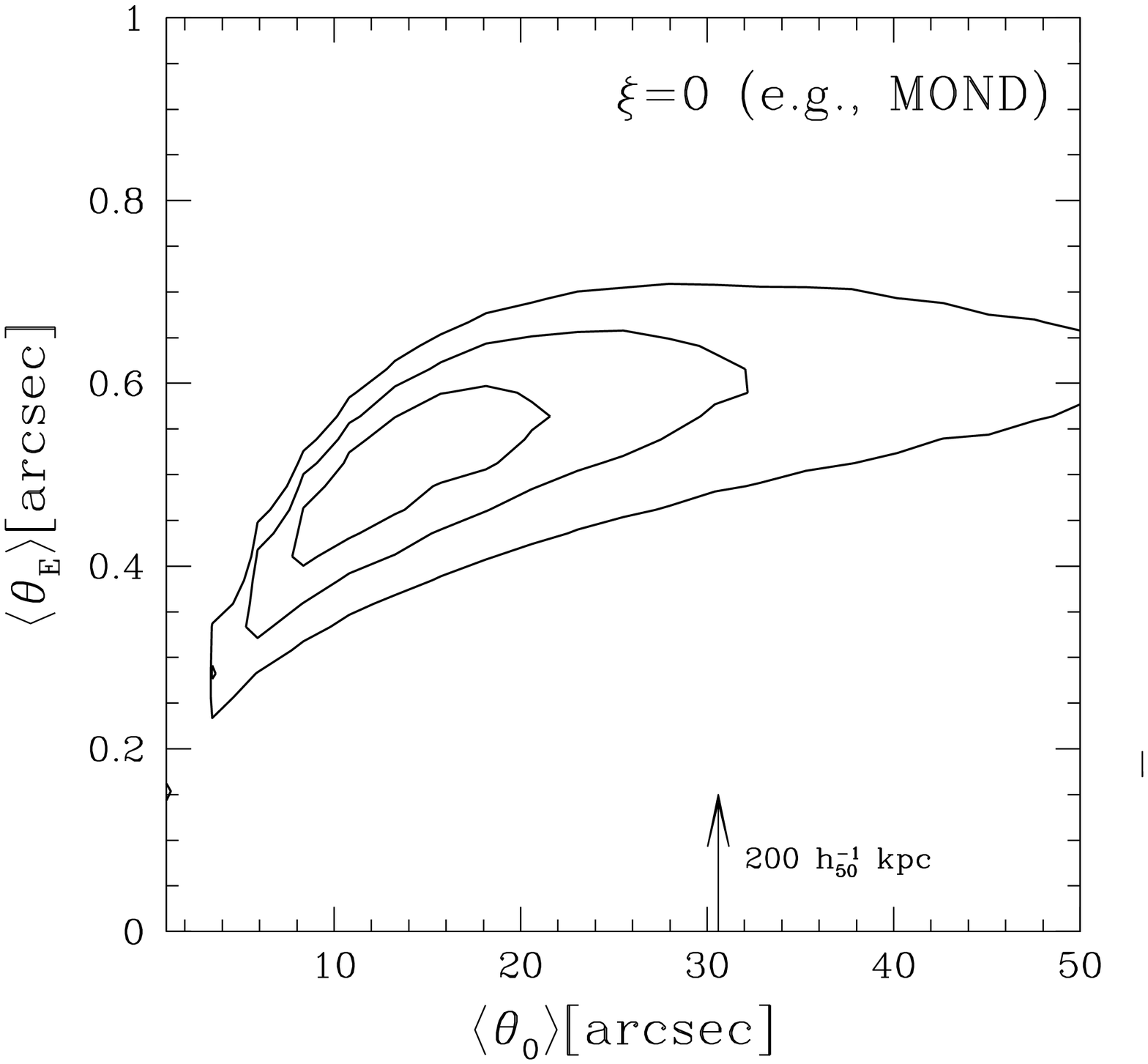}
\epsfxsize=6.8cm
\epsffile{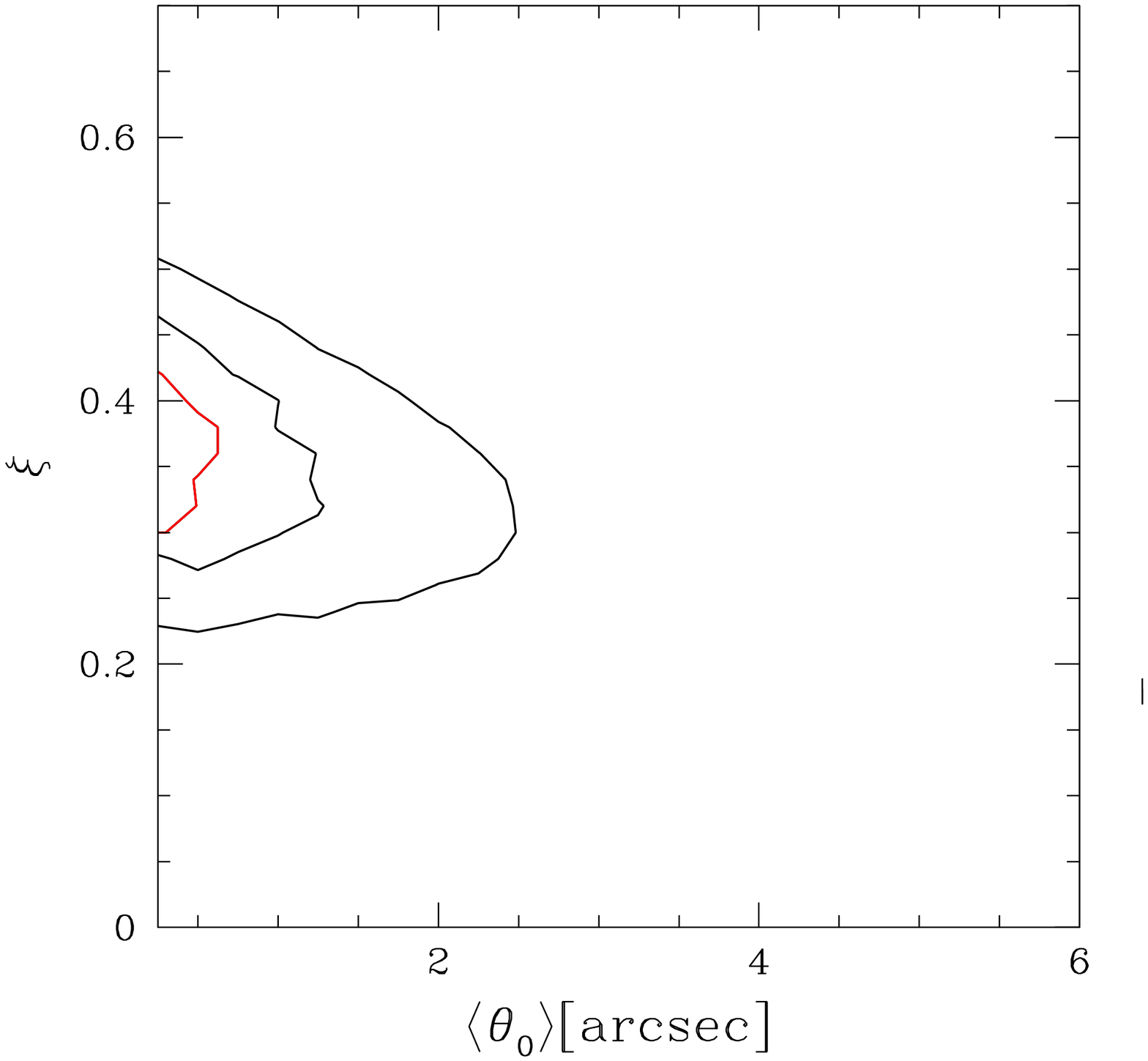}}
\caption{\footnotesize {\it Left panel:} Likelihood contours for
$\xi=0$, which can be considered as a good approximation to the MOND
case. The contours indicate the 68.3\%, 95.4\% and 99.7\% confidence
limits on two parameters jointly. The allowable range for the value of
$\theta_0$ excludes small values: the MOND regime sets in at too large
physical scales. The best fit value for the Einstein radius
corresponds to a mass-to-light ratio of $M/L_B=1.5 {\rm M}_\odot/{\rm
L}_{B\odot}$.  {\it Right panel:} The likelihood contours for the
parameters $\xi$ and $\theta_0$ as determined from a maximum
likelihood analysis of 16 deg$^2$ of RCS $R_C$ band imaging data. The
results have been marginalized over the Einstein radius $\theta_{\rm
E}$.  As expected, $\xi=1$ (GR without dark matter) is ruled out. Also
the best fit of $\xi=0.1$ from \citet{MT01a} is ruled out with high
confidence. In addition, the maximum likelihood solution is much worse
than the GR + truncated halo model presented in section~4, and the
current data are sufficient to rule out models of gravity with
deflection laws given by equation~\ref{deflaw} at the $10\sigma$
confidence level.
\label{powlaw}}
\end{center}
\end{figure}

Figure~\ref{powlaw} shows the results of a maximum likelihood analysis
of 16 deg$^2$ of $R_C$ band imaging data from the RCS.  The left panel
shows the constraints on $\theta_0$ and $\theta_{\rm E}$ for $\xi=0$
(i.e., the ``MOND'' approximation). The allowable range for the value
of $\theta_0$ excludes small values: the MOND regime sets in at too
large physical scales. The best fit value for the Einstein radius
corresponds to a mass-to-light ratio of $M/L_B=1.5 {\rm M}_\odot/{\rm
L}_{B\odot}$, which is similar to what one would expect from the stars
themselves. The right panel of Figure~\ref{powlaw} shows the
constraints on $\theta_0$ and $\xi$ when we marginalize over
$\theta_{\rm E}$.  Both the $\xi=1$ (GR without dark matter) and the
$\xi=0$ case (``MOND'') are excluded by the current data.  

Comparison of the maximum likelihood solution to the best fit ``GR +
truncated dark matter halo'' model shows that the latter is a much
better fit. The results from the ``truncated dark matter halo'' model 
indicate that the data favor a model in which the projected
mass distribution has a profile that is $\propto 1/r$ on small scales, and
steepens on larger scales. This explains why the $\xi=0$ model
results in large values of $\theta_0$, and why the data favor
a value of $\xi\sim 0.35$: the best fit model is a compromise
between the $1/r$ and $1/r^2$ profiles, reflecting the change in
slope in the density profile. Comparison of the likelihoods
of the deflection law models given by equation~\ref{deflaw} and the
``truncated halo'' model shows that the former models are excluded
at the $10\sigma$ confidence level.

The problems with the deflection laws considered by \citet{MT01a}
arise because of the form of the deflection law on large scales.  As
mentioned above, the law of gravity has to return to the $1/\theta$
form on large scales, and consequently, such theories should not
only specify a scale where the effect of alternative gravity becomes
important, but also large scale $\theta_{\rm out}$, where the theory
returns to ``normal'' gravity.

To allow for such a scale, we modify the \cite{MT01a} deflection law
(for $\xi=0$, i.e., the ``MOND'' case) to

\begin{equation}
\tilde\alpha(\theta)=-\frac{\theta_{\rm E}^2}{\theta\theta_0}
\left({\theta+\theta_0-\frac{\theta^2}{\theta+\theta_{\rm out}}}\right),
\end{equation}

which gives a shear

\begin{equation}
\tilde\gamma(\theta)=\frac{\theta_{\rm E}^2}{\theta_0\theta^2}
\left[{\frac{\theta}{2}+\theta_0-\frac{\theta^3}{2(\theta+\theta_{\rm out})^2}}\right],
\end{equation}

For $\theta\gg\theta_{\rm out}$ this modified deflection law leads to
a deflection angle (or shear) that is a factor $\theta_{\rm
out}/\theta_0$ times the GR value. Hence, the ratio $\theta_{\rm
out}/\theta_0$ is a measure of the ``mass descrepancy''. Comparison
with the truncated halo model shows that both profiles are nearly
identical on scales larger than a few $\theta_0$, provided the
truncation parameter $s=\theta_{\rm out}$. We find that such a model
provides a good fit to the data. However, we note that the introduction
of $\theta_{\rm out}$ is rather ad hoc.

The study of the radial dependence of the deflection law is
potentially useful when it is predicted completely by the theory. The
ad hoc approach used here can fit the data, and hence gives
inconclusive results.  As mentioned above, the azimuthal variation of
the lensing signal around galaxies provides a more robust method. The
detection of an anisotropic lensing signal poses a serious problem for
alternative theories of gravity.

\section{Conclusions}

Weak gravitational lensing has shown tremendous progress in the
last few years, thanks mainly to the fact that large areas of
the sky can now be observed with wide field cameras, such
as the CFH12k camera on CFHT. In this review we have highlighted
some of the most recent results.

The most recent cosmic shear studies
\citep{Bacon02,Hoekstra02b,Refregier02, vW02} show excellent
agreement.  This is remarkable, given the fact that the data have been
taken with different instruments, filters, and depths. These most
recent results demonstrate that weak lensing by large scale structure
can play an important role in the era of precision cosmology.

Weak lensing provides a direct measure of the (dark) matter
distribution on all scales, and hence it is one of the most powerful
probes of the relation between the galaxies and the underlying mass
distribution. As a result, it provides a direct way to determine the
bias parameter $b$ and the galaxy-mass cross-correlation coefficient
$r$ as a function of scale, on scales not accessible by other
techiques. The results obtained by \cite{Hoekstra02c} suggest
significant non-linear or stochastic biasing on scales $0.5-1 h^{-1}$
Mpc. These measurements provide important observational constraints on
models of galaxy formation.

The lack of visible tracers hampers our knowledge of the gravitational
potential around galaxies at large projected radii. Weak lensing has
proven to be a powerful tool to probe the extent and shapes of dark
matter halos. The maximum likelihood analysis of RCS data indicates a
clear steepening of the mass distribution on scales larger than $\sim
250h^{-1}$ kpc \citep{Hoekstra02e}. Furthermore, the data allowed
for the first detection of the flattening of dark matter halos,
with sperical halos excluded at the 99\% confidence level.

Finally weak lensing can be used to constrain alternative models of
gravity (without dark matter). Ad hoc approaches to study the radial
dependence of the deflection law remain inconclusive, mainly because
of the lack of predictions. These theories, however, predict an
essentially isotropic lensing signal around galaxies, irrespective of
the actual deflection law. The detection of an anisotropic lensing
signal around galaxies therefore poses a serious problem for theories
of alternative gravity, such as MOND.

The main results discussed in this review were obtained in the
last few years. The progress made in such a small amount of time
suggests that exciting times are ahead, when even larger data
sets become available.

HH thanks the organizers for the invitation to an interesting
meeting.

\label{}




\end{document}